\documentclass[iop, apjl]{emulateapj}
\usepackage{graphicx}
\accepted{July 19, 2010}
\begin{document}
\title{UGC8802: A Massive Disk Galaxy in Formation}

\author{Sean M.\ Moran\altaffilmark{1},
Guinevere Kauffmann\altaffilmark{2},
Timothy M.\ Heckman\altaffilmark{1},
Javier Gracia-Carpio\altaffilmark{3},
Amelie Saintonge\altaffilmark{3},
Barbara Catinella\altaffilmark{1},
Jing Wang\altaffilmark{1},
Yan-Mei Chen\altaffilmark{1},
Linda Tacconi\altaffilmark{3},
David Schiminovich\altaffilmark{4},
Pierre Cox\altaffilmark{5}, 
Riccardo Giovanelli\altaffilmark{6},
Martha Haynes\altaffilmark{6}, and
Carsten Kramer\altaffilmark{7}
}

\altaffiltext{1}{Department of Physics and Astronomy,
  The Johns Hopkins University, 3400 N.\ Charles Street, Baltimore, MD
  21218, USA}
\email{moran@pha.jhu.edu}
\altaffiltext{2}{Max Planck Institut f\"{u}r Astrophysik,
  Karl-Schwarzschild-Str. 1 , D-85741 Garching, Germany}
\altaffiltext{3}{Max Planck Institut f\"{u}r Extraterrestrische Physik, 
   Giessesbach-Str., 85748 Garching, Germany}
\altaffiltext{4}{ Department of Astronomy, Columbia University, 
550 West 120th Street, New York, New York 10027, USA} 
\altaffiltext{5}{Institut de RadioAstronomie Millimétrique ,  
300 rue de la Piscine, 38406 Saint Martin d'Hères, France}
\altaffiltext{6}{Department of Astronomy, 610 Space Sciences Building, Cornell University, Ithaca, NY 14853, USA}  
\altaffiltext{7}{Instituto de RadioAstronomía Milimétrica, Avenida Divina Pastora, 7,    
E 18012 Granada, Spain}

\begin{abstract}
We report new observations of the galaxy UGC8802 obtained through GASS, the
GALEX Arecibo SDSS Survey, which show this galaxy to be in a 
remarkable evolutionary state. UGC8802 (GASS35981) is a disk 
galaxy with stellar 
mass M$_*=2\times10^{10} $M$_\odot$ which appears to contain an
additional $2.1\times10^{10}$M$_\odot$ of HI gas. 
New millimeter observations with the IRAM
30m telescope indicate a molecular gas mass only a tenth this large.
Using  deep long-slit spectroscopy, we
examine the spatially resolved star formation rate and
metallicity profiles of GASS35981 for clues to its history. We find that
the star formation surface density in this galaxy is low
($\Sigma_{\rm SFR}=0.003 {\rm M}_{\odot}$ yr$^{-1}$ kpc$^{-2}$) and that the  star formation
is spread remarkably evenly across the galaxy. The low molecular gas masses measured in
our three IRAM pointings are largely consistent with the total star formation
measured within the same apertures.
Our MMT long-slit spectrum  reveals a sharp
drop in metallicity in the outer disk of GASS35981.
The ratio of current star formation rate to existing stellar mass
surface density in the outer disk is extremely high, implying that all the stars 
must have formed within the past $\sim$1~Gyr.  At current star formation rates, however, 
GASS35981  will not consume its HI reservoir  for another 5-7  Gyr. 
Despite its exceptionally large gas fraction for a galaxy this massive, 
GASS35981 has a regular rotation curve and exhibits  no sign of a   
recent interaction or merger. 
We speculate that GASS35981
may have  acquired its gas  directly
from the inter-galactic medium,  and that it and other similar galaxies identified in the
GASS survey may provide  rare local glimpses of gas accretion processes that were more common 
during the prime epoch of disk galaxy formation at $z\sim1$.
\end{abstract}


\section{Introduction}
One of the largest gaps in our understanding of how galaxies
form and evolve is the question of how gas---the raw material for star
formation---flows into and out of galaxies, and how these flows
regulate star formation in these systems. Although $\Lambda$CDM 
simulations make specific predictions for how  structure in the dark matter
assembles through hierarchical clustering, the assembly of the visible, baryonic
components of galaxies  is still a subject of considerable controversy.

The currently favored
theoretical picture is that cold gas flows along filaments 
into the centers of assembling dark matter halos at high redshifts ($z>2$), and this 
process can build massive  galaxies rapidly at early times (e.g., Genel et al 2008;
Dekel et al 2009). At lower redshift,
massive spiral  galaxies are thought to accrete gas much more slowly,
at a rate of less
than a few solar masses per year, from a surrounding  hot corona (Maller \& Bullock 2004;
Kauffmann et al 2006; Peek, Putman \& Sommer-Larsen 2008; Binney, Nipoti \& Fraternali 2009).
Gas may also continue to accrete through major or minor mergers with other galaxies,
although this is not believed to be the primary way in which ongoing star formation in
local spiral galaxies is fuelled at present (Sancisi et al. 2008).
Outflows, whether from star-formation or AGN-related feedback
mechanisms, are another piece of the puzzle, and their role in 
shutting down star formation,  limiting the efficiency with which stars are able to form,  
or even providing fuel for ongoing star formation in the form of recycled wind material
(Oppenheimer et al. 2009), 
is similarly debated.

In order to learn more about cold gas in nearby galaxies, we  
are carrying out the GALEX Arecibo SDSS
Survey (GASS)\footnote{http://www.mpa-garching.mpg.de/GASS} (Catinella et
al. 2010). GASS is designed
to measure the neutral hydrogen content of a representative sample of $\sim 1000$
galaxies uniformly selected from the SDSS spectroscopic and GALEX imaging
surveys, with stellar masses in the range $10^{10}-10^{11.5} {\rm M}_{\odot}$ and
redshifts in the range $0.025<z<0.05$.  
As GASS observations are designed to detect HI down to a gas-fraction
limit of 1.5\%, the full GASS sample will be the first HI survey able to
place meaningful, unbiased constraints on the atomic gas reservoirs that 
may contribute to future growth in massive galaxies. 

We are also pursuing a companion project on the IRAM 30m
telescope, COLD GASS\footnote{http://www.mpa-garching.mpg.de/COLD$\textunderscore$GASS},
which will obtain accurate and homogeneous molecular gas masses for 
a subset of $\sim 300$ galaxies from the GASS sample. These data 
will allow us to characterize the balance
between atomic and molecular gas in the galaxies in our sample, and
understand the  physical processes that determine how the condensed baryons
are partitioned into stars, HI and H$_2$ in the local Universe.

In this paper, we report on UGC8802, an extraordinary galaxy
blindly selected for inclusion in the GASS
sample (under the catalog name GASS35981, used hereafter), 
which contains a reservoir of HI $>10^{10}$~M$_\odot$,
at least equal in mass to the galaxy's entire stellar content. 
This galaxy contains
less than one tenth this mass in H$_2$ and also has  a rather modest
star formation rate (SFR). 
We describe the spectroscopic follow-up that enables us to conclude that
the outer disk of this galaxy is currently  forming from
gas that has likely accreted from the external environment.                                                                         
In the following, we adopt a standard $\Lambda$CDM cosmology with
H$_0=70$km~s$^{-1}$~Mpc$^{-1}$, $\Omega_m=0.3$ and $\Omega_\Lambda=0.7$.

\begin{figure}[t]
\centering
\includegraphics[height=2.5in, clip, trim=5.75in 0in 0in 0in]{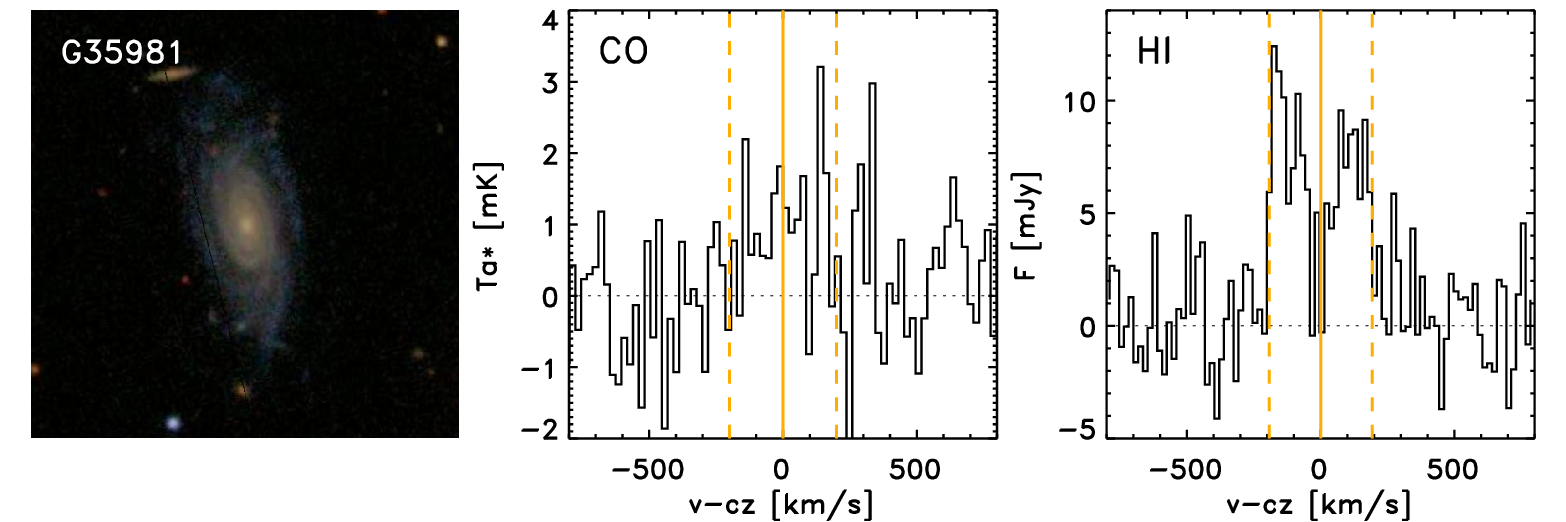}
\includegraphics[height=2.5in]{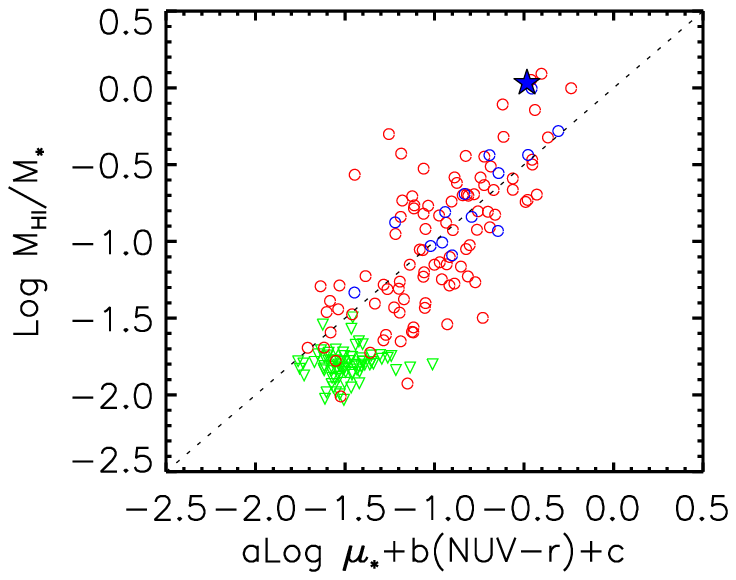}
\caption{\label{gas_prof} Top: HI spectrum of
  GASS35981 from the Cornell HI Digital Archive (Springob et al. 2005). 
 The solid orange line indicates the measured HI line
  center, and the dashed lines mark the velocity width of the
  spectrum. Bottom: GASS35981
 on the `gas fundamental plane' presented in Catinella et al. (2010),
 relating HI mass fractions to stellar mass surface density ($\mu_*$) and
 NUV--{\it r} color. The best-fit values of the coefficients {\it a},
 {\it b}, and {\it c} used in constructing the plane 
 are: $a=-0.332$, $b=-0.240$, and $c=2.856$.
 GASS35981 is marked with a blue star. Red circles indicate
 HI-detected galaxies from GASS, and green triangles mark upper limits
 on gas fraction. Blue circles denote a set of galaxies randomly drawn 
 from previous HI surveys, and added to the GASS galaxies 
 in the proper proportion to generate a statistically unbiased sample
 (see Catinella et al. (2010)).}
\end{figure}

\begin{figure*}[t]
\includegraphics[width=2\columnwidth]{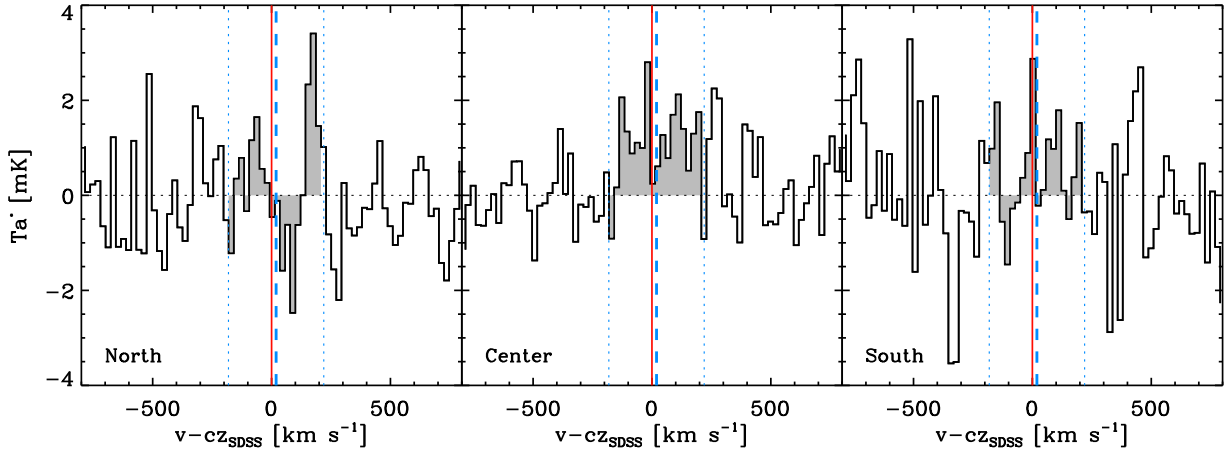}
\vspace{0.2cm}
\caption{\label{CO} CO spectra for all three IRAM pointings, as
  labeled on the figure. The red lines mark the optical
  redshift of GASS35981, the blue dashed lines show
  the velocity centroid of the central CO emission, and the dotted
  lines represent the width of the window function that was 
  used to measure the flux in the CO lines, covering an area marked by
  grey shading under the spectrum. In the central pointing, the peak
  at $\sim300$~km~s$^{-1}$ is believed to be noise, as it does not
  appear in both polarizations; it therefore is excluded from our measurement.}
\end{figure*}

\section{Observations}
GASS35981 (also SDSS~J135308.36$+$354250.5, in addition to UGC8802), 
was selected for inclusion in the
GASS parent sample because it has photometry from both  SDSS and GALEX,
is located in a region of sky accessible to Arecibo, and has
stellar mass of M$_*=2 \times 10^{10} {\rm M}_\odot$ and redshift
of $z=0.0411$ that fit into our targeted range. 
GASS35981 has  pre-existing HI observations available from the Cornell 
HI Digital Archive (Springob et al. 2005). 
The mass of HI in GASS35981 is
estimated from the line flux to be $2.1\times 10^{10}$M$_\odot$,
and the rest-frame velocity width of the HI line is  $W_{50}=360\pm25$ km~s$^{-1}$
(Figure~\ref{gas_prof}).
We display the HI archive spectrum in 
Figure~\ref{gas_prof}. We note that there is 
no evidence for contamination  from possible companion 
galaxies within the 3.5$\arcmin$ Arecibo beam: one nearby galaxy has a  spectroscopic redshift
of 0.14, and two additional faint
companions have SDSS photometric redshifts consistent with $z=0.14$.

The lower panel of Figure~\ref{gas_prof} shows that  GASS35981 (blue star) lies near the 
extreme end of HI fractions observed by GASS.  Indeed, its
gas fraction is comparable to the highest values measured for all
galaxies in this stellar mass range (e.g., Giovanelli et al. 2007). 
Indeed, even in HI-selected samples, which are biased towards objects
like GASS35981, galaxies in this stellar mass range
(M$_*>10^{10}$~M$_\odot$) with such high gas fractions are quite
uncommon (e.g., the ``HI Giants'' described by Garcia-Appadoo et al. 2009).

It is also clear that GASS35981 lies significantly 
above the best-fit `gas fundamental plane' (dotted line) 
relating stellar mass surface density, NUV--{\it r} color, and 
gas fraction, as described in Catinella et al. (2010). 
Because this galaxy was  an interesting outlier and a presumed easy target, 
GASS35981 was selected for inclusion in our initial COLD GASS
pilot program to obtain molecular gas measurements with the IRAM 30m 
telescope.

Observations of GASS35981 in the J=1--0 rotational transition of CO
were made at 3mm with the IRAM 30m telescope, in three different
pointings: one at the galaxy center, and one each to the north and south, one 
beam-width (22$\arcsec$) away along the galaxy major axis. 
Data were taken in June and August 2009, using the WILMA and
4MHz backends simultaneously to record the data, and the CLASS\footnote{http://www.iram.fr/IRAMFR/GILDAS} software to
process them. Individual scans were examined, and
a linear baseline subtracted from each of them.  After rejection of 
scans with unstable baselines due to, e.g., poor atmospheric conditions, 
the data were combined and binned to a spectral resolution of 21
km~s$^{-1}$. 

The CO line is detected in the central pointing with S/N$=5.8$, 
for an integrated line flux of $T_a^*=0.43\pm0.07$K~km~s$^{-1}$ within a
400~km~s$^{-1}$-wide window. We set upper limits at 0.29 and 0.43 K~km~s$^{-1}$ for 
the North and South offset pointing, respectively.  Adopting a 
conversion factor of $X_{CO}=4.4$M$_\odot$/L$^\prime$ (where L$^\prime$ has units of
K~km~s$^{-1}$~pc$^{-2}$), these fluxes correspond to an 
H$_2$ mass of $8.8\times10^8$ M$_\odot$ in the central 22$\arcsec$ pointing, 
and upper limits of 5.9 and $8.8\times10^8$ M$_\odot$ in the offset
pointings. Using an aperture correction for the 22$\arcsec$
beam of the IRAM telescope based on resolved CO maps of nearby 
galaxies, we estimate a total H$_2$ mass of $1.45\times10^9$ M$_\odot$ 
for the galaxy, based on the detection in the central pointing.  
Following the prescription of Springob et al. (2005), we measure a 
line width of W50$=335\pm 20$km~s$^{-1}$. This corresponds to a rest-frame
width of 321~km~s$^{-1}$. The spectra of all three pointings
are shown in Figure~\ref{CO}.

\begin{figure}[t]
\includegraphics[width=\columnwidth]{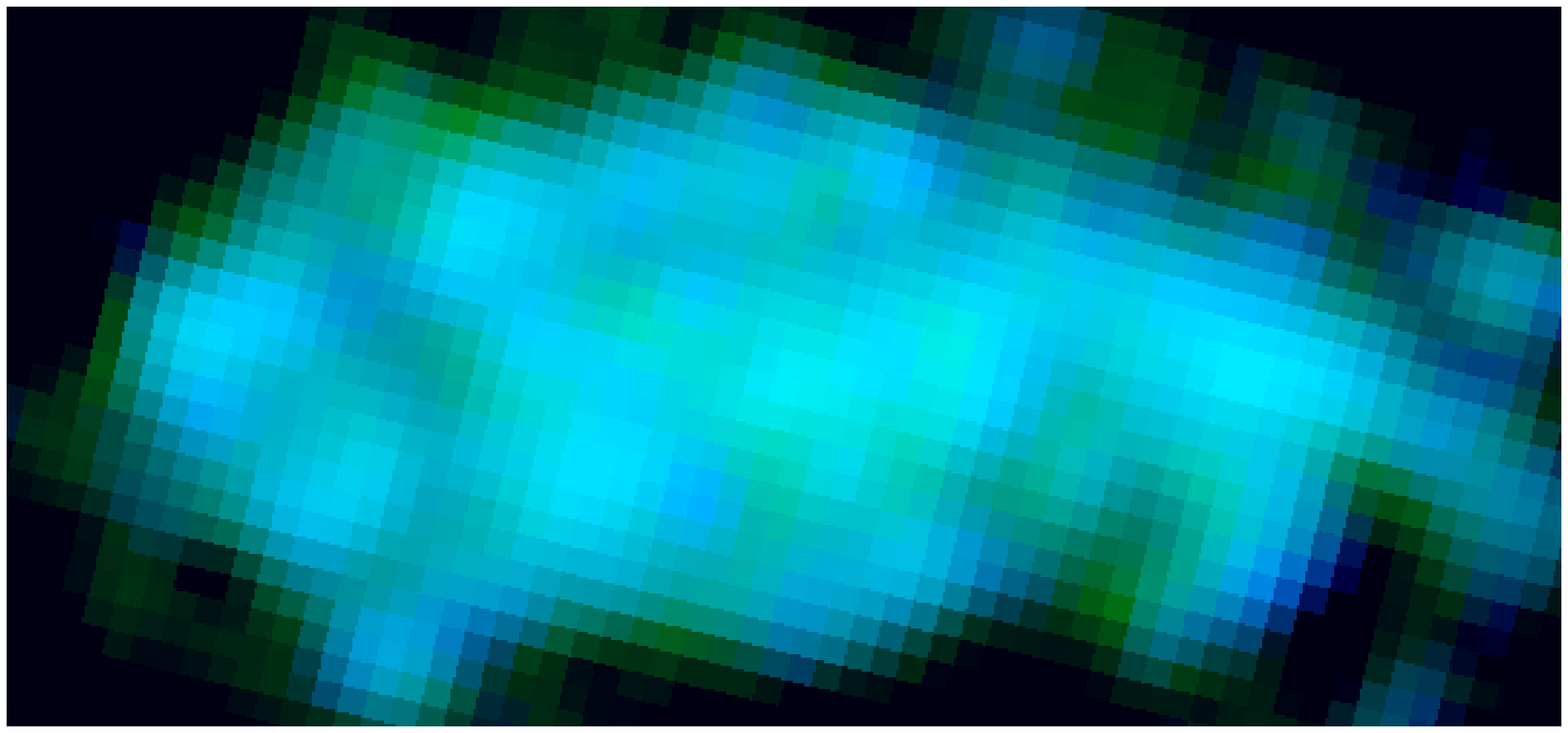}
\includegraphics[width=\columnwidth]{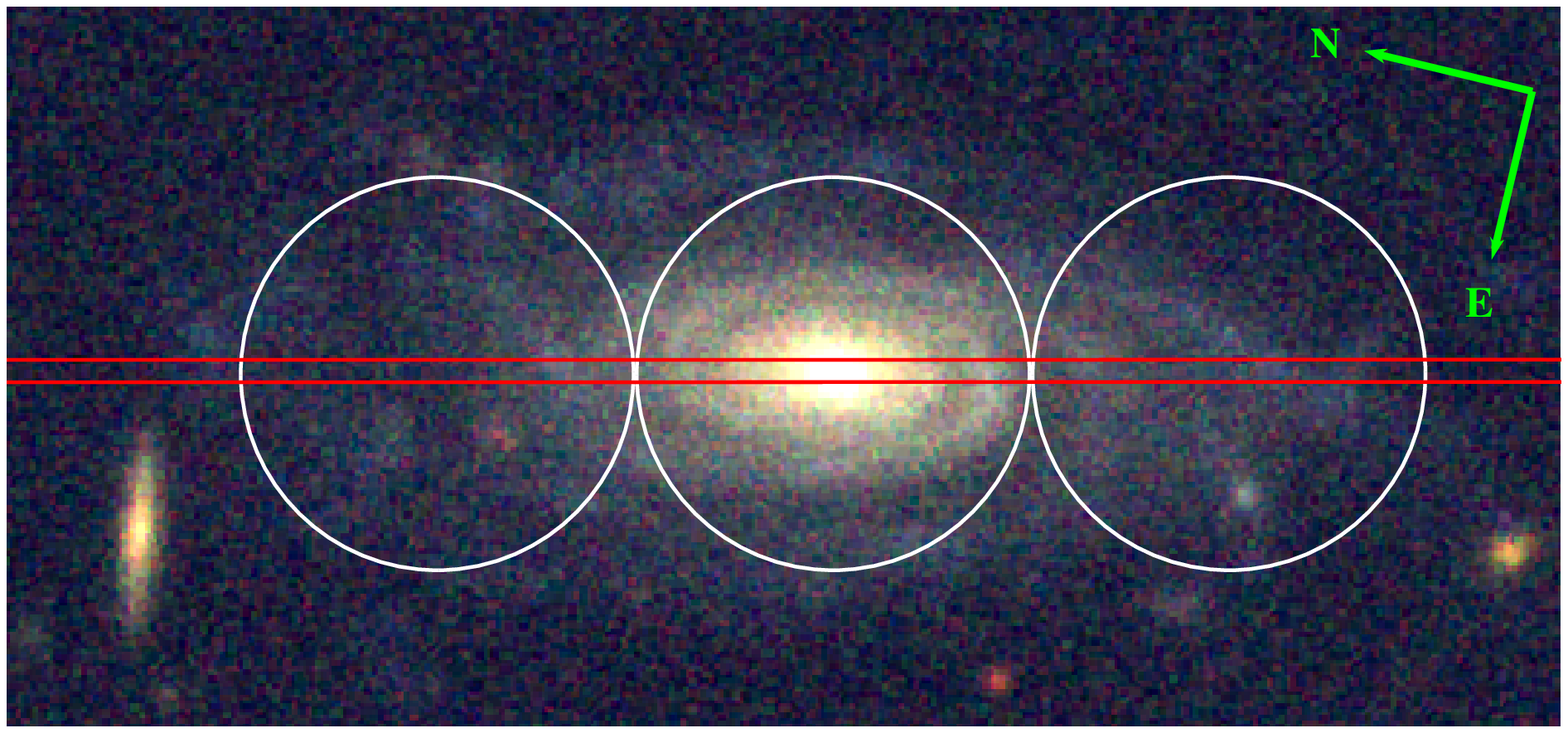}
\includegraphics[width=\columnwidth]{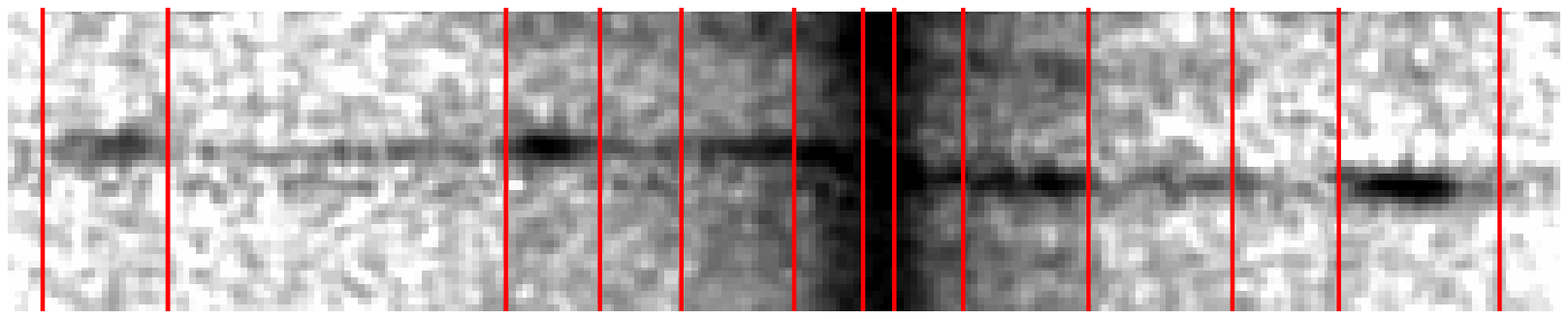}
\includegraphics[width=\columnwidth]{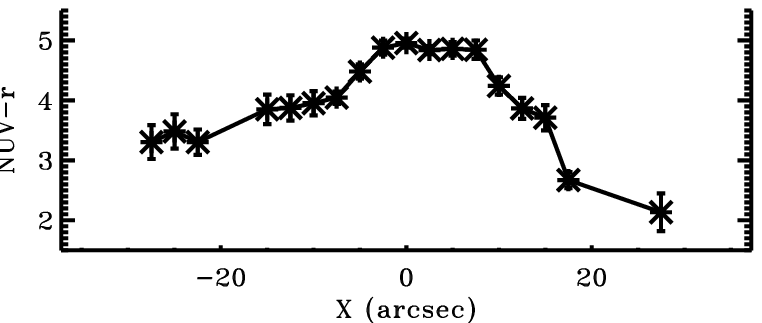}
\caption{\label{images} {\bf Top:} GALEX UV image of
  GASS35981 with FUV in blue and NUV in green, smoothed by a Gaussian
  with FWHM 3.5\arcsec, slightly smaller than the GALEX psf of
  4.5\arcsec. The dimensions of the image are  87\arcsec wide by 42\arcsec tall. {\bf Upper
    Middle:} Color SDSS image of GASS35981 constructed from {\it gri}
  images, and displayed at the same scale and size as the GALEX image
  above. Orientation and width of the spectroscopic slit are 
  overlayed in red, with north indicated by the compass rose. White
  circles overlayed indicate the three IRAM CO pointings, with circle
  diameter equal to the 22\arcsec FWHM IRAM beam size. {\bf Lower
    Middle:} A cutout of the 
  spectrum around the region of the H$\alpha$ emission line is shown
  at the same spatial scale along the x-axis. Bright 
  H$\alpha$ knots can be clearly associated with faint blue regions of the outer
  galaxy that intersect the slit. Vertical red lines mark out the
  twelve spatial regions over which we have co-added our spectrum. All
spectroscopic measurements in this paper are measured on these twelve
spectra. {\bf Bottom:} {\it NUV--r} color profile of GASS35981
extracted from a region overlapping our slit.
  Note that the X-axis
arcsecond scale applies to all four panels of this figure.} 
\end{figure}

Follow-up long-slit spectroscopy of GASS35981 was obtained on 20
November, 2009, using the Blue Channel Spectrograph on the 6.5m MMT
telescope on Mt. Hopkins, AZ. The spectrum was obtained with a slit
of width 1.25\arcsec oriented to PA$=13.4^\circ$ on the sky, such that the slit runs
along the major axis of the galaxy, as indicated in the second panel of Figure~\ref{images}. GASS35981 was observed in 2x600s
exposures with the 500 line grating. The spectrum covers a wavelength
range of $\sim3900-7000$\mbox{\AA} at a spectral resolution of $\sim4$\mbox{\AA}
FWHM, equivalent to $\sigma\sim75$km~s$^{-1}$ in the rest frame of the 
galaxy. The slit
samples the galaxy spatially with $0\farcs3$ pixels (equivalent to
0.25~kpc in physical units), over a total slit length of 150$\arcsec$, which is much larger than
the galaxy itself. 

Data were reduced in IDL with routines from the
publicly available Low-Redux
package\footnote{http://www.ucolick.org/$\sim$xavier/LowRedux/} maintained
by X. Prochaska, which itself is an adaptation of the Princeton SDSS reduction
code\footnote{http://spectro.princeton.edu} to more general long-slit 
reductions. The code performs standard biasing, flat-fielding,
cosmic-ray rejection, and sky estimation on each exposure. We then
co-add the sky-subtracted exposures through a custom-written routine
that verifies and adjusts the alignment of exposures before co-addition. 
Flux calibration was achieved via observation of the spectrophotometric standard
BD$+$17~4708. We note that we have not attempted
to apply any correction for atmospheric dispersion, even though
GASS35981 was observed at an airmass of 1.8. 
The expected atmospheric 
dispersion from blue to red end for our configuration, as estimated from 
Filippenko \& Greenstein (1984), is comparable to the size of our slit. More 
importantly, it is much smaller than the typical spatial regions we average over 
in the analysis presented below.
A cut-out of the fully-reduced spatially-resolved spectrum centered on the H$\alpha$
emission line, including only that portion of the slit where
significant flux from the galaxy is detected, is displayed in Figure~\ref{images}.

\section{Analysis}
To understand why GASS35981 contains such a high abundance of HI, but
little molecular gas, we analyze our long-slit spectrum with an
eye toward extracting information on  the dynamics of the galaxy
and the properties of its stellar population---including current star
formation rate, stellar population age and metallicity, and dust 
properties---as a function of radius across the galaxy.
We likewise re-analyze the SDSS {\it ugriz} and GALEX FUV
and NUV images of GASS35981  to 
obtain independent estimates of the  radial gradients in colors and
star formation histories, as described below.

\begin{figure}[t]
\includegraphics[width=\columnwidth]{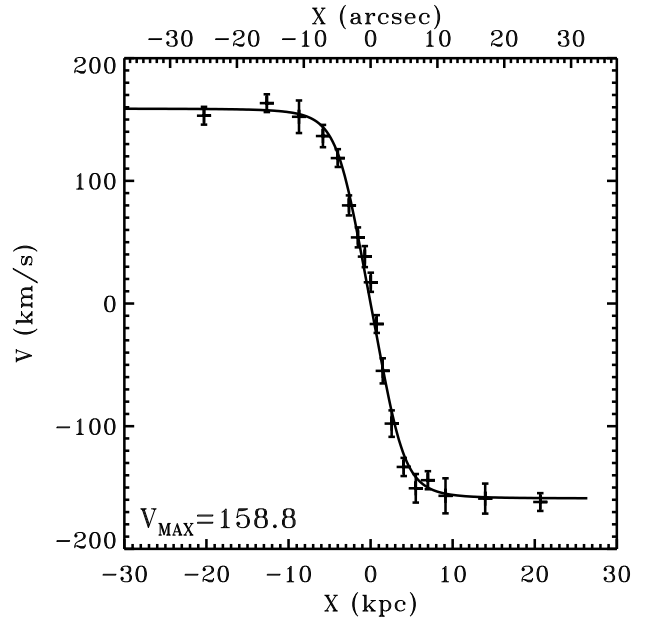}
\caption{\label{rot} Rotation curve of GASS35981 measured primarily
  from cross-correlation of the spectrum to templates, supplemented by
  measurements of the H$\alpha$ centroid for the two outermost bins to
  each side. Solid line indicates the best-fitting model of the form
  described in the text, which is symmetric to positive and negative
  positions.}
\end{figure}

\subsection{Rotation Curve}
To measure the rotation curve of GASS35981, we first bin our 
spectrum spatially to achieve a signal to noise of at least 6.0 
(\mbox{\AA}$^{-1}$) in each bin, the minimum needed for a reliable
velocity estimate. We define spatial bins by working outward from 
the galaxy center (where S/N is high even
without binning), co-adding rows one at a time until the
minimum S/N is reached, or until adding rows fails to increase the 
S/N. Since strong H$\alpha$ emission is detected in GASS35981 
out to much larger radius than the continuum flux (see
Figure~\ref{images}), we also include
several additional bins that correspond to knots of bright H$\alpha$ 
emission in the spectrum. For each binned spectrum, we determine the
effective spatial position by calculating the luminosity-weighted
average radius of all spatial positions that entered the co-add.

We then measure the radial velocity as a
function of radius in two ways: 1)  by cross-correlating each
spectrum against galaxy templates to determine the velocity, 2) by
fitting directly for the centroids of the H$\alpha$ emission line. The
method utilizing H$\alpha$ allows us to extend our rotation
curves to larger radii.  The two methods yield consistent curves
at radii where we can fit templates and measure emission line centroids.
The resulting rotation
curve is plotted in Figure~\ref{rot}, not corrected in any way 
for the inclination of the galaxy. The solid line plotted over the
measured curve is the best-fitting rotation curve of the
form $V(R)=V_{\rm MAX} R/(R^a+R_s^a)^{1/a}+\Delta V$ (B\o hm et al. 2004;
Moran et al. 2007), where $R$ is the
radius, $a$ and $R_s$ are free parameters that govern the shape
of the rotation curve and its turn-over, and $\Delta V$ is the offset
of the galaxy's central velocity from the  redshift obtained
from the SDSS spectrum (also left as a free parameter). We measure a
circular velocity for GASS35981 of $2V_{\rm MAX}=318 \pm 10$~km~s$^{-1}$.
This is consistent with the value obtained from the CO spectrum 
($W=321\pm 10$~km~s$^{-1}$). 

Table~1 summarizes some of the basic properties of GASS35981, including the
masses and line widths of the various components.
The agreement between CO and H$\alpha$ widths is not surprising 
because at the radius of the IRAM beam (11 arcsec), the optical
velocity has already reached its maximum value. 
It is interesting, however, that 
the rest-frame HI velocity width, $W=360\pm25$~km~s$^{-1}$,
appears to be somewhat larger than the estimates  from both  CO and H$\alpha$. 
\footnote {We note that the velocity widths are not deprojected to edge-on 
(the galaxy's inclination is 63.25 degrees from face-on, estimated 
as $\cos (i)=b/a=0.45$).}
Since the optical rotation curve of GASS35981 is flat, 
one might have expected a better agreement between the HI and H$\alpha$
widths. 

\begin{figure*}[t]
\includegraphics[width=2\columnwidth]{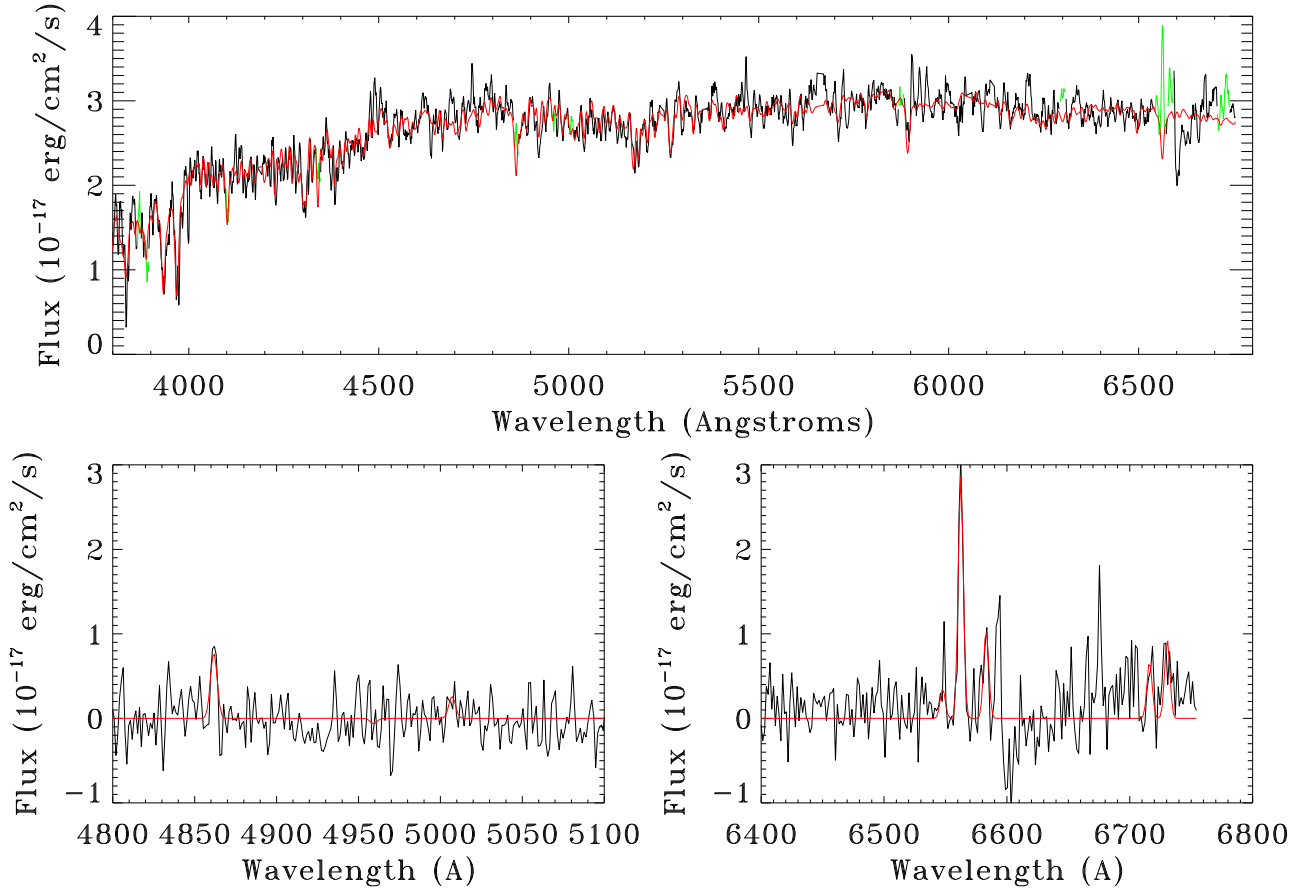}
\caption{\label{fits}Example of absorption template (top) and emission line (bottom)
  fits to the observed spectrum of GASS35981, here shown for a
  spectrum extracted 2--3kpc from the center of the galaxy. In the top
panel, the spectrum is plotted in black with regions that were masked
for the fitting indicated in green. The best-fitting model spectrum is
over-plotted in red. In the bottom two panels, two cutouts of the
spectrum near the H$\beta$ and H$\alpha$ emission lines are replotted
with the best-fit model subtracted off. Our Gaussian fits to key
emission lines are then over-plotted in red.}
\end{figure*}

Even though the significance of the  difference  is  
marginal (smaller than 2$\sigma$), it is interesting to comment on 
possible causes of the effect.
Based on a statistical analysis of a large data set with both H$\alpha$ and
HI spectroscopy, Catinella et al. (2007) find a
systematic difference between optical and HI velocities of normal spiral galaxies
with flat rotation curves of a factor 1.06 (see their figure 6).
After applying a 6.5 km~s$^{-1}$ turbulent motion correction (as in 
Catinella et al. 2007), and accounting for such a scaling factor, 
the HI width of GASS35981 is 333 km~s$^{-1}$, which is in slightly  better agreement with
the optical rotation curve and CO linewidth. 
The residual difference might be explained by 
uncertain statistical and
turbulent motion corrections, or it might provide evidence for accretion of HI at large
radii. It is interesting that the  HI profile shown in Figure~\ref{gas_prof} is noticeably asymmetric.
We will come back to this point in the final discussion. 

\subsection{Spatial Binning of the Spectrum}
We now correct each row in
the unbinned 2D spectrum to a common rest-frame 
using the rotation curve in Figure~\ref{rot}. This procedure is necessary if we wish to obtain
accurate measurements of spectral lines.  We then re-bin the spectrum  
to a slightly higher targeted S/N  of $\sim 8$.
To each side, the two bins at highest radii are chosen by hand, because there is no clear 
detection of the stellar continuum this far out. As can be seen in
Figure~\ref{images}, nebular emission is still clearly detected in these
regions, so we choose the binning to correspond to clear 
transitions between the bright or faint knots of H$\alpha$. 
We end up with a total of 12 radial bins, and we mark
the limits of  each  with red lines on the
spectrum in Figure~\ref{images}.
We use this re-binned spectrum in our subsequent analysis. 

\subsection{Continuum \& Emission Line Fitting}

We employ a modified version of the technique of Brinchmann et al (2004)
to measure the strengths of key emission and absorption lines as a
function of radius across GASS35981. 
The spectrum from each of our 12 spatial positions is first fit to a
linear combination of templates drawn from Bruzual \& Charlot (2003) single
stellar population models, masking regions of the spectrum occupied by
common emission lines. 
Best-fit model spectra are generated
individually for both a low- and high-metallicity set of Bruzual \&
Charlot models (0.02Z$_\odot$ and Z$_\odot$, respectively), and we
retain the one with lowest reduced $\chi^2$ of the fit. Near the
center of the galaxy, these fits unambiguously favor the
solar-metallicity model.  Low-metallicity templates provide a
slightly better fit in each of the two outermost bins where we can
measure the stellar continuum ( $R\sim15$~kpc on each side). However, we
caution that the S/N in the continuum is $<5$ here, and so the
best-fitting metallicity is unlikely to be a reliable indicator. 
The results of an example fit  are shown in Figure~\ref{fits}.

Next, we subtract the
best-fitting stellar continuum model from the measured spectrum,
creating an emission-line only spectrum where the Balmer emission lines can
be measured free of contamination from the underlying stellar absorption.
In the two outermost bins, 
only nebular emission is detected and  no continuum fitting
is possible. In these cases, we fit a low-order polynomial to the
spectrum to correct for small imperfections in our sky
subtraction, which arise  when co-adding across a large
portion of the slit. We then measure emission lines using the  
polynomial-subtracted spectrum. 

\begin{deluxetable}{llll}
\tablecolumns{4}
\small
\tablewidth{0pt}
\tablecaption{GASS35981 Key Properties}
\startdata
\hline \hline \\[-2ex]
M$_{*}$ & $2.0\pm0.3\times10^{10} \textrm{M}_\odot$ & W$_{H\alpha}$\tablenotemark{b} & $318\pm10$~km~s$^{-1}$ \\
M$_{HI}\tablenotemark{a}$ & $2.1\pm0.4\times10^{10} \textrm{M}_\odot$ & W$_{HI}$\tablenotemark{a,b} & $360\pm25$~km~s$^{-1}$\\
M$_{CO}$ & $1.4\pm0.2\times10^{9} \textrm{M}_\odot$ & W$_{CO}$\tablenotemark{b} & $321\pm19$~km~s$^{-1}$\\
$M_r$ & -21.2 & $M_{NUV}$ & -18.7 \\
R$_{25}$\tablenotemark{c} & 24~kpc & R$_e$\tablenotemark{d} & 9.8~kpc\\
$z$ & 0.0411 & SFR$_{H\alpha}$
& $3.7\pm0.3$ M$_\odot$
\enddata
\tablenotetext{a}{Springob et al. (2005)}
\tablenotetext{b}{Rest-frame line width}
\tablenotetext{c}{SDSS {\it r}-band isophotal major axis radius (25mag$/\arcsec^2$)}
\tablenotetext{d}{SDSS {\it r}-band Petrosian half-light radius}
\end{deluxetable}

We fit a Gaussian function to the emission lines, with the width of the Gaussian constrained
to a single value for all lines in a given spectrum. The  
positions of the line centroids  are constrained to their rest-wavelengths.  The
only free parameters in the fit are the amplitude of each line and an overall
velocity offset term. For each
line, we also recalculate the continuum level in a small region within
50\mbox{\AA} of each fitted line, and subtract this small residual
from the template fits to further refine the measured emission line flux.
In addition to  the Balmer lines H$\alpha$, H$\beta$, and H$\gamma$, used to
estimate dust extinction and star formation rate (see below), we also
measure the forbidden lines [OIII]~5007, [NII]~6548 and 6584, and 
[SII]~6717 and 6731 to measure metallicity across the galaxy.

\begin{figure*}[t]
\includegraphics[width=2\columnwidth]{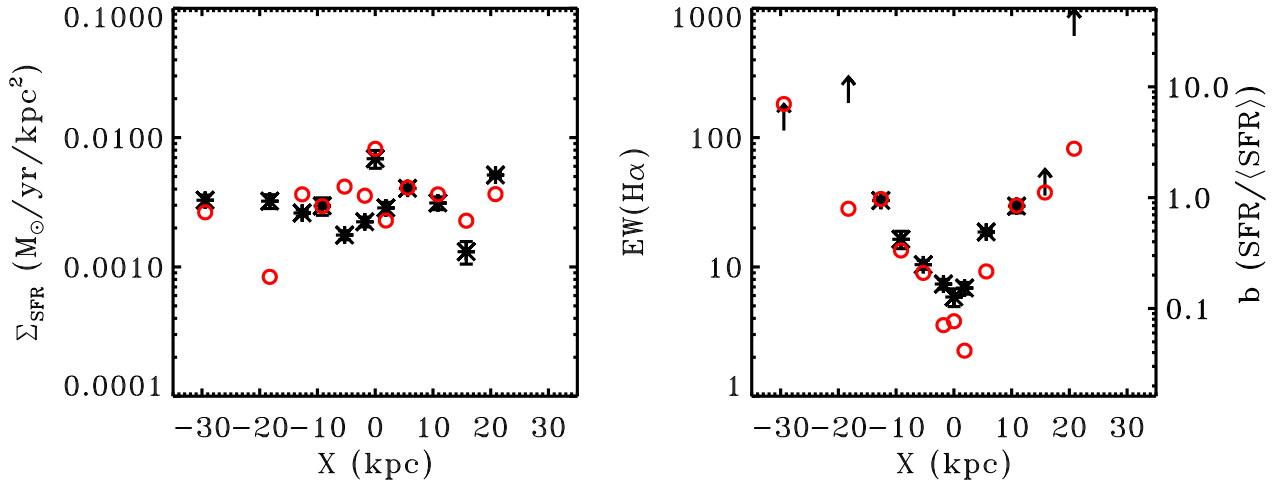}
\caption{\label{sfr} Left: Star formation rate surface density, in M$_\odot$~yr$^{-1}$~kpc$^{-2}$, as a
  function of radius for each spatial position across
  GASS35981. Black points indicate SFR surface density measured directly from
  dust-corrected H$\alpha$ luminosity, while red circles are estimated
  from fits to the photometry as described in the text. Current star
  formation appears to be evenly spread across the
  galaxy, with a scatter of only 0.2 dex. Right:
  EW(H$\alpha$) as a function of position (black points and lower
  limits), which can be expressed as the
  ratio of current to past-averaged star formation, {\it b}, as
  described in the text. Red circles indicate {\it b} values estimated
  directly from the photometric fits. }
\end{figure*}

We limit our analysis of stellar absorption features to the 
$D4000_n$ index, which measures the
strength of the 4000\mbox{\AA} break. This index is  an indicator of stellar
population age (see, e.g., Kauffmann et al. 2003). We use $D4000_n$
values  measured from the best-fitting absorption line templates,
rather than from  the spectrum itself. The two agree well for spectral  bins with
high S/N.  We choose  the model measurement as we expect it
to yield a more reliable estimate in  bins where S/N in the continuum is
only $\sim3-5$, because the model fit  utilizes 
information from the  entire spectrum. 

We estimate the dust extinction within the nebular gas by calculating the
Balmer decrement, which we  define as  the ratio of H$\alpha$/H$\beta$ to the
case B recombination ratio of 2.87 (Osterbrock 1989). We adopt the 
formula $A_V=1.9655R_V \log({\rm H}\alpha/{\rm H}\beta/2.87)$, where we assume
$R_V=3.1$ and adopt the Calzetti (2001) extinction curve. 
We further refine the estimates of $A_V$ by comparing to the
H$\gamma$ line, which is only strong enough to be measured in emission
in a few cases. However, even as an upper limit, H$\gamma$   provides a useful
constraint. If, after correcting all three lines for dust based on our
original $A_V$ estimate, the ratios H$\gamma$/H$\alpha$ and/or
H$\gamma$/H$\beta$ are {\it inconsistent} with the expected  values of
0.474 and 0.166, respectively, we adjust $A_V$ upward or
downward---staying within the $1\sigma$ errors on the original 
measurement---to improve the agreement with these values  as
far as possible.

We note that all SFRs reported below have been calculated after
correcting fluxes for extinction using the Balmer
decrement. Likewise, our equivalent widths have been corrected for
dust using our measured Balmer decrements and assuming 
E(B-V) of the stellar light is $\sim0.44$ E(B-V) of the gas, as in 
Calzetti (2001). 

\subsection{Colors and Star Formation Rates from the Photometry}
We measure the  UV/optical color profiles of GASS35981 using SDSS and GALEX photometry
in two ways. First, we measure fluxes in a  
5$\arcsec$ wide strip running along the major axis of the galaxy.
Our chosen aperture lies on top of, but is wider than the spectroscopic slit.
To ensure that the fluxes measured using the  SDSS images are well-matched
to those measured from the GALEX images,
we convolve the  SDSS images with a Gaussian of FWHM 4.5\arcsec (the size of the GALEX PSF).
The resulting color profile is displayed in the bottom panel of  Figure~\ref{images}. 

For more precise comparison to the spectroscopy (but at the expense of
lower S/N in  lower surface brightness regions), we also measured colors 
through apertures with widths identical to the slit segments 
containing each of our twelve spectral bins. These apertures are smaller
than the GALEX PSF, so we scale the SDSS {\it u}-band flux measured
in the smaller apertures by the ratio of  FUV or NUV to {\it u} measured in the
wider aperture at the same position along the slit. We use these measurements to      
estimate SFR and stellar mass surface densities by  
fitting the full UV through
optical SEDs to population synthesis models as described in Wang et al. (2010, in preparation).   
Our  method is very  similar to that described in Salim et al. (2007). 

To optimize the fits, we constrain the                       
internal extinction of the galaxies in the library of models to be consistent
with the measured  Balmer-decrement values for this galaxy.
Assuming that the attenuation of the starlight  is 0.44 times
the extinction measured in  nebular gas (Calzetti 2001),   
we obtain  $\tau_v=0.5$  averaged over our 12 spectral bins, with a dispersion 
of  $\pm0.25$.  We use this as a prior in our parameter estimation.  

\section{Results}
In this section, we analyse  
star formation rates, stellar population ages and metallicities of the nebular
gas  as a function of position in 
GASS35981. We  show that star formation is spread fairly
evenly across the disk of the galaxy, but the ratio of 
current star formation to  stellar mass surface density in the outskirts of the galaxy
is very high.
This suggests the outer disk has  been formed recently. 
We then show that the modest  H$_2$ content implied by the CO
observations matches expectations from the star formation rate profile of the galaxy. 
We also  examine the metallicity profiles,  
finding a sharp drop in metallicity that corroborates our conclusion that stars in the  outer
disk were formed recently.

\subsection{Star Formation in GASS35981}
After correcting H$\alpha$ luminosities for dust, we measure star
formation rates using  the equation in  Meurer et
al. (2009): SFR ($M_\odot$~yr$^{-1}$)=L$_{{\rm H}\alpha}/$($6.93\times
10^{33}$~W), corrected to a Kroupa (2001) IMF. 
Then, by dividing by the area of the galaxy
covered by each  portion of the slit ($1.25\arcsec \times
0.3\arcsec N$, where N is the number of individual rows that went into
each co-added spectrum), we estimate  the SFR surface density as a function of
position across GASS35981. This is shown in Figure~\ref{sfr}, left, 
from which it is clear that the  SFR surface density is remarkably uniform
across the galaxy. 
This result  is confirmed by the SFR
surface densities estimated from the photometry alone (\S~3.3), over-plotted as red
circles on Figure~\ref{sfr}, which match the values estimated using the spectroscopy
very well. The star formation rate surface density profile of GASS35981 is quite
different to those of ``normal'' spiral galaxies of the same stellar mass.
Figure~2 of Bigiel et al. (2008) shows that in  spirals where
H$_2$ dominates in the central regions,  $\Sigma_{\rm SFR}$ decreases from
0.01-0.1 $M_{\odot}$ yr$^{-1}$ kpc$^{-2}$ near the galaxy centers to 
less than 0.001 $M_{\odot}$ yr$^{-1}$ kpc$^{-2}$ near the edges of the optical
disk. The SFR surface density profile of GASS35981 is much  more similar to those of
HI-dominated dwarf irregular galaxies (Figure~3 of Bigiel et al.), but
these galaxies have considerably smaller stellar masses.  

\begin{figure}[t]
\includegraphics[width=\columnwidth]{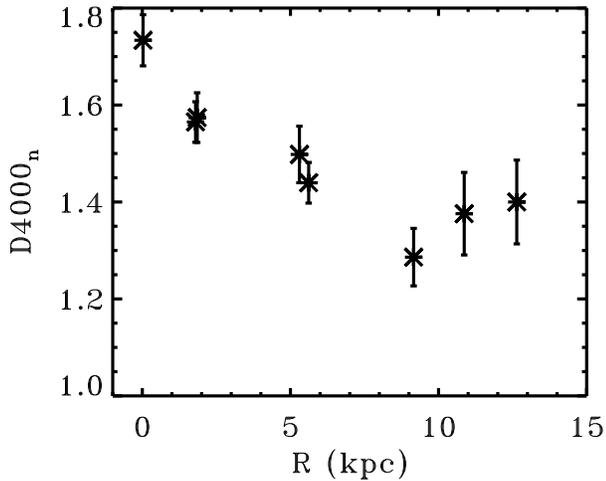}
\caption{\label{d4000} D4000$_n$ vs radius, with the index estimated
  from the best-fitting continuum model of the observed spectrum. 
In a simple single stellar population model from Bruzual \& Charlot
(2003), $D4000_n=1.3$ indicates a stellar population age of 500~Myr, while $D4000_n=1.7$  occurs at an age of $\sim3$~Gyr.}
\end{figure}

In the right panel of Figure~\ref{sfr},  
we plot  EW(H$\alpha$) as a function of position along the slit.
Since the equivalent width is the ratio of emission line flux to the underlying
continuum, it is an excellent proxy for ratio of current star
formation to past-averaged star formation (i.e., pre-existing stellar
mass), conventionally referred to as the $b$-parameter.  
The exact relation between the two depends on the past star formation
history of the galaxy. An empirical relation between the two quantities can be derived   
for galaxies in the SDSS using the star formation rates, stellar masses and
H$\alpha$ equivalent widths published in the
SDSS MPA/JHU value added 
catalogs\footnote{http://www.mpa-garching.mpg.de/SDSS/}. The
$b$-parameter is estimated from the SFRs and stellar masses using the
formulation described in Brinchmann et al. (2004). 
We use this relation to transform between EQW(H$\alpha$), which is plotted on the
left axis of  Figure~\ref{sfr}, to $b$, which is plotted on the right axis
of the same figure.  

Figure~\ref{sfr} shows that the timescale for building 
up the current stellar mass surface density at the current star formation rate
decreases strongly as a function of radius. At radii $R>10$~kpc
we find that  $b>1$, 
indicating that the SFR is currently higher than it has been over most
of the galaxy's history (Rocha-Pinto et al. 2000). 
{\it b} reaches values in excess of 10 in the very outer
regions of the galaxy, indicating that all the stars in the outer
disk could have been formed over a timescale as small as a  
few hundred Myrs. 

At large radii, the H$\alpha$ emission line is very strong, but there is                             
no detection of the underlying
continuum, yielding only a lower limit on the EW. We also 
estimated {\it b} directly from our photometric fits, 
and these values are over-plotted as red circles in
Figure~\ref{sfr}. Overall, the results agree well, though there are
two points where the estimated $b$ is significantly lower than the
limits implied by the spectroscopy. Reasons for the discrepancy could
include small errors in sky subtraction on the spectrum leading us to
underestimate the 3-$\sigma$ limit on the continuum flux, which would
subsequently cause an overestimate of {\it b}. Alternatively, since the
discrepant points are in regions where {\it b} appears to have a
strong spatial gradient, it may be that seeing and/or resolution
effects are limiting how well we can match spatial extraction regions
between spectrum and photometry, and the discrepancy simply reflects
that spatial variation. In either case, both the spectrum and
photometry agree that $b\ge1$ all across the outer disk.
 
Another possible worry is that {\it b} parameters and star
formation rates measured through  small apertures are not
representative of the galaxy as a whole; however, we obtain nearly
identical trends with radius when we estimate {\it b} and 
$\Sigma_{\rm  SFR}$ from
photometry that has been azimuthally-averaged in elliptical annuli
across the whole galaxy.

To test whether the age of the stellar 
population in the outer disk is  consistent with
the  short formation timescale deduced from the H$\alpha$   equivalent
widths, we examine the radial trends in stellar population age as
measured by the $D4000_n$ index (Figure~\ref{d4000}). Although $D4000_n$
is known to vary with metallicity as well as age (Kauffmann et
al. 2003), the metallicity of
GASS35981 appears to be fairly uniform within the restricted range of
radii over which we measure it (with only one point in
Figure~\ref{d4000} at significantly
sub-solar metallicity; see \S~4.2), and so for this radial
regime $D4000_n$ is the highest-S/N proxy for stellar population age
that we can measure. 

We see in
Figure~\ref{d4000} that $D4000_n$
appears to decrease from $\sim1.7$ in the center of the galaxy to
$1.3-1.4$ beyond $\sim10$~kpc. For a simple single stellar population (SSP)
model from 
Bruzual \& Charlot (2003), $D4000_n=1.3$ indicates a stellar population age 
of 500~Myr, while $D4000_n=1.7$  occurs at an age of $\sim3$~Gyr.
These characteristic timescales grow larger for more realistic
extended star formation histories; this will be discussed further 
in  \S~5.
The S/N of our spectrum
is not high enough to measure $D4000_n$ at very large radii, but the
fact that such young ages are implied for $R=10$~kpc is 
suggestive that the outer disk may be even younger.

\subsubsection{Analysis of the CO Measurements}
Given the order of magnitude difference between the measured HI mass
and the measured molecular gas mass, it is important to determine
whether the latter is consistent with the derived star formation rate. 
 Bigiel et al. (2008) show that  SFR surface density is
directly proportional to H$_2$ mass surface density according to the relation:
$\Sigma_{\rm SFR}=10^{-3.1} \Sigma_{\rm H_2}$ (in units of
M$_\odot$~yr$^{-1}$~kpc$^{-2}$ and  M$_\odot$~pc$^{-2}$, respectively). 
This relation holds over a large range in $\Sigma_{\rm H_2}$ 
(3-50 $M_{\odot}$ yr$^{-1}$ pc$^{-2}$) and does not appear to vary
as a function of any larger scale property of the galaxies
in the sample.

When comparing SFR and H$_2$ mass for GASS35981, we must carefully 
account for possible  aperture effects, because the
SFR surface density is measured through a narrow slit, while the H$_2$ masses are 
estimated in  three much larger apertures with FWHM  of  22\arcsec.
However, we have seen that in GASS35981 the SFR surface density 
appears to be quite uniform
across the whole galactic disk.
The CO line flux 
measured in our central pointing  corresponds to an average H$_2$
surface density across
the beam of $\Sigma_{\rm H_2}=3.6\pm0.1 M_\odot$~pc$^{-2}$, under our
assumed conversion factor $X_{CO}=4.4$M$_\odot$/L$^\prime$. 
Adopting the Bigiel et al.
relation between $\Sigma_{\rm H_2}$ and $\Sigma_{\rm SFR}$  yields a predicted 
SFR surface density $\Sigma_{\rm SFR}=0.0029\pm0.0001 M_\odot$~yr$^{-1}$~kpc$^{-2}$,
which is in excellent agreement with the observed $\Sigma_{\rm SFR}$ in this galaxy.

In the two offset
pointings where we do not detect any CO, our most stringent upper
limit (in the Northern pointing) implies that $\Sigma_{\rm H_2}<2.4
M_\odot$~pc$^{-2}$, and hence $\Sigma_{\rm SFR}<0.0019$. This is 50\%
lower than the median $\Sigma_{\rm SFR}$ we measure from our spectra,
though from Figure~\ref{sfr} we see that our photometric estimates of
$\Sigma_{\rm SFR}$  exhibit a dip in the region covered by the
northern pointing, which could explain the low  H$_2$ mass.

Alternatively, variation in the conversion factor $X_{CO}$ could
explain the discrepancy; since below we will present evidence that
the metallicity is low in the outer regions of GASS35981 (which
implies a lower amount of CO per unit of H$_2$), this
possibility seems plausible. We note that Bigiel et al. (2008)
neglected variations in $X_{CO}$ in deriving their SFR--H$_2$ relation,
and so the $\sim0.3$~dex scatter they measure likely includes a
contribution from variations in metallicity/$X_{CO}$. Though the
single discrepant point here is too little to draw conclusions from,
in a future paper we will examine possible variations 
in $X_{CO}$ across the COLD GASS sample.

\subsubsection{Total Star Formation Rate}
Finally, we estimate a total star formation rate for GASS35981 by
extrapolating  the SFR surface density measured for each spectral bin along  the slit to 
the  annulus that spans the same radial
range as that portion of the slit, and that extends half-way around the
galaxy, where it meets the corresponding annulus extrapolated from the
other side of our slit. The aperture corrections are large, but it is instructive
to compare the resulting  total, dust-corrected SFR with our purely photometric estimate.
We find that SFR$_{\rm H\alpha}=3.7\pm0.3 M_\odot$~yr$^{-1}$, which is
slightly higher than the value of $3.0\pm0.4 M_\odot$~yr$^{-1}$  estimated  from the integrated
photometry  
using  the method described in Schiminovich et al. (2010).
We find no evidence for any additional
obscured component of star formation in GASS35981, as there is no
counterpart in the Faint
Source Catalog of IRAS (Moshir et al. 1992), to a limit of 0.2Jy 
at 60$\mu$m, which corresponds to a luminosity of $L_{60}\sim10^{10}$L$_\odot$.

\subsection{Metallicity Gradient}

\begin{figure}[t]
\includegraphics[width=\columnwidth]{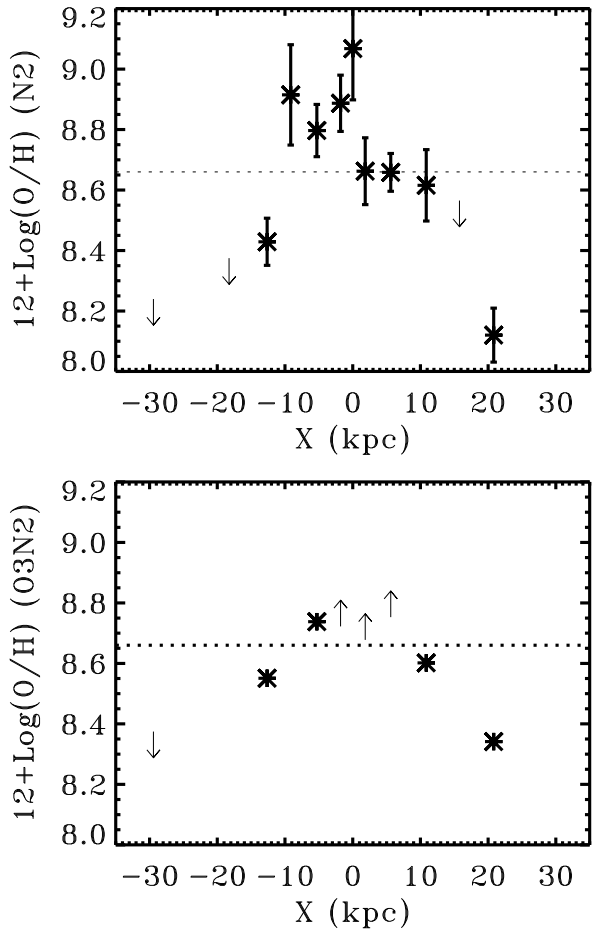}
\caption{\label{metal}{\bf Top:} Metallicity of
GASS35981 as a function of position along the slit, estimated from the
[NII]/H$\alpha$ emission line ratio (Pettini \& Pagel 2004). Solar
metallicity is marked with the dotted line. {\bf Bottom:} 
Similar metallicity estimates including the ratio of [OIII]/H$\beta$,
also from  Pettini \& Pagel, as described in the text.}
\end{figure}

We utilize the
[NII] 6584 to H$\alpha$ emission line ratio as our primary
metallicity indicator. We estimate metallicity  from the relation of
Pettini \& Pagel (2004): $12 + \log(O/H)=9.37+2.03\times N2 + 1.26\times
N2^2+0.32\times N2^3$ where $N2=\log([{\rm NII}] 6584/{\rm H}\alpha)$. The
resulting metallicities are plotted  as a function of position across GASS35981 
in Figure~\ref{metal}. In the bulge and inner disk of the galaxy
($R<15$~kpc), GASS35981  exhibits near-solar
metallicity. However, at large radii, we see  a
sharp drop in metallicity of $\sim0.5$ dex, at a radius corresponding to the start
of the blue, faint outer disk of the galaxy (Figure~\ref{images}). This is also the
region of the galaxy where the $b$ parameter exceeds unity. 

Sharp breaks to low metallicities are not common in the
portion of the larger GASS sample with follow-up MMT spectroscopy
that has been  analyzed so far. We find an average  
gradient in $\log(O/H)$ of $-0.14$~dex/$r_e$, compared to $-0.41$~dex/$r_e$ for GASS35981. 
Less than  10\% of the galaxies in our sample  exhibit metallicity below 
12+log(O/H)$<8.5$, and in most of these cases, the low metallicities are seen over the whole galaxy. 
Only $\sim15$\% of the 62 galaxies analyzed so far have  metallicity gradients
as strong as GASS35981, and even fewer of these exhibit  a sharp
drop. We note that there is one well-known local galaxy
with a metallicity gradient similar to GASS35981: M101 (van Zee et
al. 1998). The importance of this similarity is unclear, however,
since the two galaxies differ in other respects (e.g., the
distribution of star formation; Kuntz \& Snowden 2010).

As a check on our [NII]/H$\alpha$ metallicity 
estimates, we also compute  metallicity using  
$O3N2=\log({\rm [OIII]~5007/H}\beta /(\rm{[NII]~6584/H}\alpha))$  
(Pettini \& Pagel 2004), which is plotted in the bottom panel of Figure~\ref{metal}. Though intrinsically more accurate, this 
indicator provides somewhat poorer constraints for our galaxy, because
H$\beta$ is not always detected at $>3\sigma$ in our spectra. 
We note, however,  that metallicities measured in this way are entirely
consistent with those from  [NII]/H$\alpha$  , as can be seen by comparing
the two panels in Figure~\ref{metal}. 
We also use  ([SII]~$6717+{\rm [SII]}~6731$)/H$\alpha$ as an additional check, as this line depends
on metallicity in much the same way as [NII] (Dopita et al. 2006). 
Although it is detected with lower S/N in our spectrum, and is more affected by
sky subtraction residuals due to its redder wavelength, the
[SII]/H$\alpha$ ratio exhibits a similar drop at large radii. This indicates 
that peculiar Nitrogen abundances are not responsible for the effect that we see.

\begin{figure}[t]
\includegraphics[width=\columnwidth]{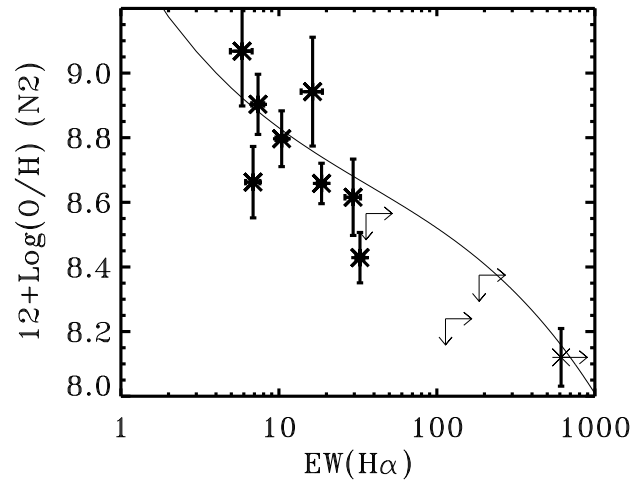}
\caption{\label{ew_metal} Metallicity from the [NII]/H$\alpha$ ratio
  as in Figure~\ref{metal}, but plotted as a function of
  EW(H$\alpha$). Increasing H$\alpha$ EW is strongly correlated with
  decreasing metallicity in GASS35981. Solid line
  indicates the relation followed by normal star-forming galaxies drawn from SDSS,
  as described in the text.}
\end{figure}

Finally,  in Figure~\ref{ew_metal} we plot 
metallicity as a function of EW(H$\alpha$).  
There is clearly a  strong correlation: 
regions of the galaxy with the highest ratio of  current to past-averaged
star formation rate are also the most metal-poor.
We also over-plot the best-fit relation between
EW(H$\alpha$) and [NII]-based metallicity for star-forming galaxies
drawn from the SDSS MPA/JHU value-added catalog. These measurements apply
to the region of the galaxies sampled by the 3\arcsec -diameter SDSS fiber.  
The relation
between star formation rate and metallicity measured within GASS35981
follows that exhibited by the general population of star-forming galaxies in SDSS.
We note, however, that the galaxies with high EW(H$\alpha$) and low metallicities
are generally much lower mass systems.

\section{Discussion}
We have seen that GASS35981 is currently undergoing star
formation that is  evenly spread across the galaxy. 
In the outskirts,
however, a sharp drop in metallicity and a correspondingly large
increase in the EW(H$\alpha$) suggest that the bulk of the stellar
mass in this region has been formed only recently, and out of a low
metallicity reservoir of gas.

In this section, we ask how long ago the gas began forming its stars, and we consider
the processes that may have delivered the gas to the galaxy.

\subsection{A Model Star-formation History}
We construct a simple toy model of the galaxy's
star formation history (SFH), in  the bulge ($R<5$~kpc), inner
disk ($5<R<15$~kpc) and outer disk ($R>15$~kpc)  of
the galaxy. Using  constraints derived from the galaxy's
current star formation rate and stellar mass surface density, we  build a
simple model that is consistent with observations in all three
locations. 

Our observations show that the current star-formation is
uniformly distributed across the galaxy, so we postulate that the 
overall SFH can be modeled by  a period of constant star formation of
some length that is the same everywhere, superimposed on  an
older stellar population that dominates at the  center and is virtually
absent at large  radii.

Our model incorporates the following constraints:
\begin{itemize}
\item  At $R>15$~kpc, the metallicity is low and the 
  {\it b} parameter is extremely high. Let us assume that this indicates 
  {\it all} of the existing stellar mass has been
  built during  the current star-formation episode. For the three
  outer spectral bins  with the lowest metallicities, the timescales for
  forming the entire stellar population at the currently observed star formation
  rate range from
  0.7--2.0 Gyr. We thus adopt 1~Gyr as our reference value. 
\item We then subtract the stellar mass formed in the current episode from 
 the observed stellar mass in the bulge and the inner disk, and assign
 the resulting mass to the old stellar population.
  In the inner disk, the mass surface density in the old stellar component  is $\sim10$
  times that in the outer disk, while in  the bulge region, this factor is
  closer to  $60$.
  For simplicity, we will concentrate all of the
  star formation needed to build this much mass
  into a single episode of 1~Gyr length ending at $z=1$ (7~Gyr in the past).

\end{itemize}

\begin{figure}
\includegraphics[width=\columnwidth]{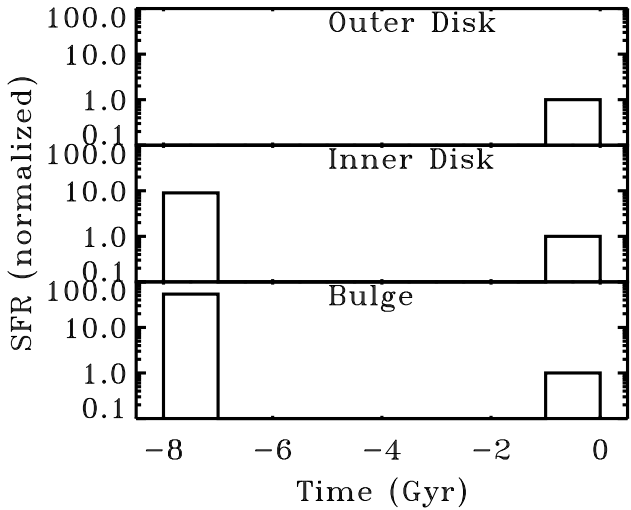}
\caption{\label{sfh} Schematic of a model star formation history that
  replicates key features of GASS35981's stellar population in each of
three radial regimes. We model the current star formation as a single
episode of length 1~Gyr proceeding at the same rate everywhere across
the galaxy (normalized to $SFR=1$ in this diagram). The
underlying old stellar population is modeled as another 1~Gyr-long burst occurring
at $z=1$, with amplitude  constrained by the observed  present-day stellar
mass surface density in each radial regime.}
\end{figure}
\vspace{0.4cm}

\par \noindent A simple schematic of this model SFH is shown in
Figure~\ref{sfh}. We then use  Bruzual \& Charlot (2003) models
to predict $D4000_n$ in the inner disk and bulge, obtaining values
of 1.3 and 1.5, respectively. 
These values are in good agreement with the measured values of 
1.4 and 1.6 in the corresponding radial bins. Thus, through only a
simple partition of mass into old and young components, we can
reasonably reproduce one of the key spectral features of GASS35981.
We note that varying the formation time of 
the old component or doubling its timescale to 2~Gyr changes the results very little.
 Likewise, 
adjusting the length of the more recent star-forming episode within
the range 0.7--2~Gyr does not significantly affect our results. 

Finally, we can estimate the total fraction of the stellar mass added to the 
galaxy in the most recent star formation episode and find that it is
around 20\%. Note, however, that at its current star formation rate, GASS35981 will
not exhaust its HI reservoir for another 5--7 Gyr.

\subsection{ Was the Gas Captured from Another Galaxy?}

We have concluded that a new episode of star formation began
approximately 1~Gyr ago in GASS35981. It is tempting to speculate that
renewed star formation in this galaxy is connected to the acquisition of its
large HI reservoir.  
 
So how did  GASS35981 acquire its gas? One possibility is that the gas 
was accreted as the result of an interaction with another gas-rich system.  
GASS35981 does have two luminous neighbors at projected
distances of 300~kpc and 400~kpc with 
redshift differences of $<400$km~s$^{-1}$---i.e., GASS35981
does appear to be a member of a loose group.  
We do not, however, see any signs of an interaction or merger in 
either the rotation curve or optical morphology of GASS35981.

Furthermore, a scenario in which GASS35981 is left with
$2 \times 10^{10}$ M$_\odot$ of HI after an  encounter with
either of its neighbors seems fairly implausible. 
Simulations predict that no more than 20\%
of the gas  mass of any donor galaxy should  
be stripped in an encounter (Bournaud, private communication), so we would  
require a donor galaxy with $>10^{11}$ M$_\odot$ of HI! Such an
encounter {\it could}, however, be responsible
for jostling an already-present HI reservoir out 
of equilibrium, causing it to form stars.                 

If the gas was not acquired from another passing galaxy, one might speculate that GASS35981 
accreted its HI reservoir directly from the surrounding
intergalactic medium at some point in the past. Since blind HI surveys such as ALFALFA
(Giovanelli et al. 2005) and HIPASS (Barnes et al. 2001) do not find  dark HI clouds 
of $10^{10}$M$_\odot$, the gas must have entered GASS35981 from an
ionized phase if it entered all at once or over a short
timescale. Since its two identified neighbors suggest that GASS35981
resides in a group, the presence of intra-group gas could be 
fueling an unusually high accretion rate onto GASS35981. 
Indeed, accretion from the intra-group medium has been 
speculated to be responsible for a number of other peculiar 
star-forming systems (e.g., Beaulieu et al. 2010).

Alternatively, the gas
reservoir could have been built slowly through multiple accretions of
smaller gas clouds or streams, which could be either neutral or ionized. Under
this scenario, star formation would need to be suppressed somehow 
during the buildup of the reservoir. The galaxy formation models of 
Birnboim et al. (2007) exhibit quiescent, reservoir building periods
similar to what would be needed here, but in general they
apply to somewhat higher mass galaxies than GASS35981, and also may
not be valid at $z\sim0$. Multiple minor mergers with gas-rich
dwarfs could also supply the gas, but it becomes even harder in this
case to imagine how the HI could build up over time rather than
form stars with each new accretion event.

\section{Conclusions}

We have reported on the remarkable galaxy GASS35981, 
 a disk galaxy with stellar mass M$_*=2\times10^{10}$~M$_\odot$ which
contains an additional $2.1\times10^{10}$~M$_\odot$ of HI gas. 
Millimeter observations indicate a molecular gas mass only a tenth this
high.
Through follow-up long-slit spectroscopy, plus SED fitting using  our
UV through optical photometry we have shown that:

\begin{itemize}
\item Star formation is  evenly spread across the galaxy,
  at a surface density of $\Sigma_{\rm SFR}=0.003$~M$_\odot$~yr$^{-1}$~kpc$^{-2}$.
\item  The proportion of the  stellar mass contributed by  
  the current star-formation episode rises towards the outer regions of
  the  galaxy, reaching a peak at $R=30$~kpc, where
  the entire disk must have formed in the past Gyr.                           
\item Interstellar  metallicities exhibit a sharp
  drop at $R>15$ kpc, coincident with a sharp rise in EW(H$\alpha$).
  This is consistent with recent infall of  lower-metallicity gas (Tinsley \& Larson 1978). 
\item The H$\alpha$ rotation curve is  regular and symmetric, and 
  reveals no signs of a recent interaction or merger that could have
  deposited the gas and/or triggered the recent star formation episode.    
\end{itemize}

The main conclusion from our observations is that GASS35981 appears to
be in the early stages of formation of its outer stellar disk.

We are not able to provide conclusive answers to questions pertaining to the
origin and fate of the gas in this galaxy with this data set alone.
Scenarios in which the gas was acquired in a recent  merging event are
disfavoured because of the extremely regular kinematics of the disk. The
HI mass of GASS35981 is too large to be easily explained by gas transfer from a passing
galaxy. We therefore speculate that GASS35981 acquired its gas directly
from the intergalactic medium.  Although our observations show that the
stars in the outer disk formed within the last Gyr, this does not mean
that the gas was also acquired less than  
1 Gyr ago. It is also unclear whether GASS35981 
will continue forming stars in its current low-efficiency state,
or whether the gas will flow inwards towards the bulge, and GASS35981
will eventually develop into a more normal massive spiral galaxy with a star
formation surface density that decreases as a function of radius.

Questions concerning the eventual fate of the gas
can be addressed using the larger samples that will
be provided by the full GASS and COLD GASS surveys in the future. By studying trends
in SFR surface density, mean stellar age, metallicity, and stellar mass
profiles as a function of  atomic and molecular gas content for complete
samples of galaxies, we  hope to map out evolutionary sequences
in disk galaxy formation.

Answers to questions concerning the origin of the gas will likely require
a different approach.  Our comparison of the HI linewidth of GASS35981
with its CO line width and  H$\alpha$ rotation curve 
yield tantalizing hints that the
atomic gas  may not be in  equilibrium with the rest of the
galaxy. In addition, the HI spectrum in Figure~\ref{gas_prof} is clearly asymmetric
about the line center.  High resolution HI mapping of GASS35981 will be
needed to understand the dynamical state of the gas in more detail. Even
so, such observations are unlikely to prove that the HI originated from
a more diffuse (and unseen) reservoir of IGM gas. This can only be done
if we are able to find  tracers of this gas, for example absorption lines
in the spectra of background quasars that arise when the quasar light
passes through the circumgalactic medium of the galaxy (Cen \& Ostriker 1999). These tracers
must then be linked with galaxies like GASS35981.

\acknowledgments The authors thank J. Brinchmann and C. Tremonti for making
available their code for analysis of spectra. Based on observations
carried out with the IRAM 30m telescope. IRAM is supported by 
INSU/CNRS (France), MPG (Germany) and IGN (Spain). The Arecibo
Observatory is part of the National Astronomy and Ionosphere Center,
which is operated by Cornell University under a cooperative agreement
with the National Science Foundation.

Observations 
reported here were obtained in part at the MMT Observatory, a facility
operated jointly by the Smithsonian Institution and the University of
Arizona. MMT telescope time was granted by NOAO, through the Telescope 
System Instrumentation Program (TSIP). TSIP is funded by NSF.

Funding for the SDSS has been provided by the Alfred P.
Sloan Foundation, the Participating Institutions, the National Science
Foundation, the U.S. Department of Energy, the National Aeronautics
and Space Administration, the Japanese Monbukagakusho, the Max Planck
Society, and the Higher Education Funding Council for England.

\end{document}